# Quark Mass Renormalization on the Lattice with Staggered Fermions

Weonjong Lee

*Department of Physics*
*Pupin Physics Laboratories*
*Columbia University*
*New York, N.Y. 10027, U.S.A.*

Oct. 6, 1993

## Abstract

The QCD light quark mass renormalized at a 1 GeV scale in the $\overline{MS}$ scheme is obtained from the numerical results of the lattice QCD simulation with staggered fermions. The primary emphasis is given to the connection between the lattice and continuum parameters. The results are compared with those from the QCD sum rule.

# 1 Introduction

A connection between lattice and continuum renormalized parameters is necessary to compare the lattice observables with the continuum observables (or experimental data) and to check whether the numerical simulation of lattice QCD makes physical sense. The bridge for the QCD coupling between the lattice and the continuum was done through the weak coupling expansion to one-loop order[1, 2, 3, 4]. The bridge for the QCD quark mass with 4-flavor staggered dynamical fermions was done in Ref. [5, 6].

Building on that work, we will attempt to relate the bare, light quark mass of 1.6 MeV obtained from 2-flavor staggered fermion simulations[9, 8, 7] with the renormalized light quark mass of between 3.5 and 9 MeV defined at a 1 GeV scale in the $\overline{MS}$ scheme and deduced from the QCD sum rules[10].

In order to make this comparison we must assume that the coupling constant used in the lattice calculation ($\beta = 6/g^2 = 5.7$) is within the perturbative region[11, 12].

We use and compare a number of possible perturbative approaches including the mean field method (tadpole improvement) suggested by Lepage and Mackenzie.

# 2 Lattice QCD with Staggered Fermions

Lattice QCD has a lot of difficulties in implementing quark flavor dynamics. There are two popular methods to put the fermions on the lattice: one is the Wilson fermion formalism and the other the staggered fermion formalism[5]. In the limit of zero quark mass, the staggered fermion scheme has remnants of chiral symmetry and may be preferred for the numerical calculation of the meson and hadron spectrum.

## 2.1 Lattice QCD Action

The current numerical simulation of lattice QCD is based on the following action [7, 13, 14].

$$S = -\sum_{x,\mu} \frac{1}{2}\eta_\mu(x) \left[\bar{\chi}(x)U_\mu^\dagger(x)\chi(x+a_\mu) - \bar{\chi}(x+a_\mu)U_\mu(x)\chi(x)\right]$$

$$-ma \sum_x \bar{\chi}(x)\chi(x) + S_{gluon}$$
$$= -\bar{X}^t (\slashed{D} + ma) X + S_{gluon} \tag{1}$$

where

$$S_{gluon} = \sum_\Box \beta \left[1 - (\frac{1}{N_c}) \text{Re Tr } U_\Box\right]$$

and the definition of $\eta_\mu(x)$ is given in Ref. [5, 15]. The current approach to flavor dynamics is as follows:

$$\langle O \rangle = \frac{\int d[U]\ O\ \exp(-S_{gluon}) \left[\det\left(\slashed{D} + ma\right)\right]^{\frac{N_f}{4}}}{\int d[U] \exp(-S_{gluon}) \left[\det\left(\slashed{D} + ma\right)\right]^{\frac{N_f}{4}}} \tag{2}$$

for $N_f$ degenerate flavors of mass m. It is important to notice that $m$ is the bare quark mass and an input parametric mass for the numerical simulation. We will let $m = M_0(a)$ represent the physical value for this bare lattice quark mass for a given value of the lattice spacing $a$.

## 2.2  Bare Quark Mass on the Lattice

From the numerical results [8], at $\beta = 5.7$,

$$\begin{aligned}
M_\pi^2 a^2 &= -0.011(3) + 6.48(14) ma \quad (\chi^2 = 4.2) \\
M_\rho a &= 0.31(2) + 9.7(8) ma \quad (\chi^2 = 0.06) \\
M_N a &= 0.47(3) + 15.0(1.2) ma \quad (\chi^2 = 2.6) \\
f_\pi a &= 0.044(1)
\end{aligned} \tag{3}$$

We can choose our conventions in the following way. The lattice QCD scale is obtained such that $M_\rho$ has the physical value in the limit of $m \to 0$. The bare quark mass, $M_0(a)$ is obtained such that the ratio, $\frac{M_\pi}{M_\rho}$ has the physical value in the limit of $m \to M_0(a)$. At $\beta = 5.7$,

$$\begin{aligned}
\frac{1}{a} &= \frac{M_\rho}{0.31} = 2.48 GeV \\
M_0 &= \frac{0.31^2}{6.48\ a} \frac{M_\pi^2}{M_\rho^2} = 1.2 MeV
\end{aligned} \tag{4}$$

Since $M_N/M_\rho = 1.52$ in the calculation of Ref.[8], we will get different values of $1/a$ and $M_0(a)$ if we choose $M_N$ to set the scale;

$$\frac{1}{a} = \frac{M_N}{0.47} = 2.00 GeV$$
$$M_0 = \frac{0.47^2}{6.48} \frac{M_\pi^2}{a \, M_N^2} = 1.51 MeV \quad (5)$$

We may also choose $f_\pi$ to set the scale $1/a$.

$$\frac{1}{a} = \frac{f_\pi}{0.044} = 2.12 GeV \quad (6)$$

## 3  Mass Renormalization in the Continuum

The most common renormlization scheme in continuum QCD is the $\overline{MS}$ scheme in Feynmann gauge. Using dimensional regularization, the one-loop self-energy contribution can be calculated as follows:

$$\Sigma(p, M_0, g_0, \epsilon) = g_0^2 R(N_c) \int \frac{d^d k}{(2\pi)^d} \Gamma_\mu \frac{1}{i(\slashed{p}+\slashed{k}) + M_0} \Gamma_\nu \frac{\delta_{\mu\nu}}{k^2}$$
$$= i\slashed{p}\Sigma_1 + M_0 \Sigma_2 \quad (7)$$
$$\text{and } d = 4 - \epsilon \; ,$$

where $M_0$ and $g_0$ are the bare quark mass and coupling respectively. The renormalized couplings are related as follows:

$$g_0 == \left(\mu_{\overline{ms}} \frac{\exp(\gamma_E/2)}{\sqrt{4\pi}}\right)^{\epsilon/2} Z_g^{\overline{ms}} \, g_{\overline{ms}}$$
$$Z_g^{\overline{ms}} = 1 - \frac{g_{\overline{ms}}^2}{16\pi^2} \frac{2}{\epsilon} \left(\frac{11}{2} - \frac{1}{3} N_f\right)$$

The quantities in Eq.(7) are given by

$$R(N_c) = \frac{N_c^2 - 1}{2N_c}$$
$$\Sigma_1 = \frac{g_{\overline{ms}}^2}{8\pi^2} R(N_c) \left(\frac{1}{\epsilon} - \int_0^1 dx (1-x) \ln[\frac{\Delta}{\mu_{\overline{ms}}^2}] - \frac{1}{2}\right) \quad (8)$$

$$\Sigma_2 = \frac{g_{\overline{ms}}^2}{8\pi^2} R(N_c) \left( \frac{4}{\epsilon} - 2\int_0^1 dx \ln[\frac{\Delta}{\mu_{\overline{ms}}^2}] - 1 \right) \tag{9}$$

$$\Delta = x(1-x)p^2 + xM_0^2 . \tag{10}$$

The renormalized quark propagator, $G$ is related to the bare quark propagator, $G_0$ as follows:

$$G^{-1}(p, M_{\overline{ms}}, g_{\overline{ms}}, \mu_{\overline{ms}})$$
$$= Z_\psi^{\overline{ms}} G_0^{-1}(p, M_0, g_0, \epsilon) |_{M_0 = Z_m^{\overline{ms}} M_{\overline{ms}} \; , \; g_0 = \mu_{\overline{ms}}^{\epsilon/2} Z_g^{\overline{ms}} g_{\overline{ms}}} \tag{11}$$

$$\text{where} \quad G_0^{-1}(p, M_0, g_0, \epsilon) = i\not{p} + M_0 + \Sigma(p, M_0, g_0, \epsilon) \tag{12}$$

In the $\overline{MS}$ scheme[18], the renormalization constants $Z_\psi^{\overline{ms}}$ and $Z_m^{\overline{ms}}$ are given to the lowest order in $g_{\overline{ms}}^2$ by

$$Z_\psi^{\overline{ms}} = 1 - \frac{g_{\overline{ms}}^2}{8\pi^2} R(N_c) \frac{1}{\epsilon}$$

$$Z_m^{\overline{ms}} = 1 - \frac{g_{\overline{ms}}^2}{8\pi^2} R(N_c) \frac{3}{\epsilon} ,$$

so the renormalized quark propagator for $N_c = 3$ is

$$G^{-1}(p, M_{\overline{ms}}, g_{\overline{ms}}, \mu_{\overline{ms}}) = \left( 1 - \frac{g_{\overline{ms}}^2}{6\pi^2} \left[ \int_0^1 dx(1-x) \ln(\frac{\Delta}{\mu_{\overline{ms}}^2}) - \frac{1}{2} \right] \right)$$
$$\cdot \left( i\not{p} + M_{\overline{ms}} \left[ 1 - \frac{g_{\overline{ms}}^2}{6\pi^2} \left( \int_0^1 dx(1+x) \ln(\frac{\Delta}{\mu_{\overline{ms}}^2}) + \frac{1}{2} \right) \right] \right) \tag{13}$$

In perturbation theory, we interpret the location of the pole in the renormalized quark-propagator as the physical mass of the particle, regardless of the renormalization scheme[20]. So if we define the pole location as $p^2 = -M_{phy}^2$, $M_{phy}$ should satisfy the following condition.

$$M_{phy} = M_{\overline{ms}} \left[ 1 - \frac{g_{\overline{ms}}^2}{6\pi^2} \left( \int_0^1 dx(1+x) \ln(\frac{\Delta}{\mu_{\overline{ms}}^2}) + \frac{1}{2} \right) \right]_{p^2 = -M_{phy}^2} \tag{14}$$

Since the pole location is independent of the renormalization scheme, it will be important for connecting the continuum and lattice schemes below.

# 4  Mass Renormalization on the Lattice

Since the ultra-violet divergences are already regularized by the lattice, only a subtraction prescription is needed for renormalization. One of the most important characteristics common to both MS and $\overline{MS}$ schemes is that the renormalization constants are independent of any dimensionful lagrangian parameters such as the mass[19]. We will follow a similar procedure in defining a renormalization scheme on the lattice, choosing renormalization constants in the lattice-regularization formalism that are independent of any dimensionful lagrangian parameters such as the mass.

## 4.1  Renormalization Prescription

In order to implement the minimal subtraction idea in the lattice regularized theory[21, 6], only divergent terms such as $(\ln[a\mu_L] - \text{Constant})^i$ with $i \leq n$ will be subtracted consistently in the lattice regularized Feynmann diagrams of $g_L^{2n}$ order. The arbitrary constant will be chosen in a physically reasonable way, as is described in detail later. In such a minimally-subtracted theory, it is well known that the renormalization group functions, $\beta(g)$ and $\gamma_m(g)$ are independent of the covariant gauge choice[16].

Let the renormalization constants, $Z_g^L$, $Z_m^L$ and $Z_\psi^L$ be defined as follows [6]:

$$g_0(a) = Z_g^L(t, g_L(\mu_L))g_L(\mu_L) \tag{15}$$
$$M_0(a) = Z_m^L(t, g_L(\mu_L))M_L(\mu_L) \tag{16}$$
$$\psi_0(a) = \sqrt{Z_\psi^L(t, g_L(\mu_L))}\psi_L(\mu_L) \tag{17}$$

where $t = -2\ln(a\mu_L)$ and $\mu_L$ is the lattice renormalization scale introduced by our subtraction.

For the coupling constant renormalization, we choose a prescription so that $\Lambda_L$, given by the conventional formula [1, 2, 4]

$$\Lambda_L^2 = \frac{1}{a^2} \exp\left(-\frac{1}{\beta_0 g_0(a)^2}\right) [\,\beta_0 g_0(a)^2\,]^{-\frac{\beta_1}{\beta_0^2}} \{\, 1 + O(g_0(a)^2)\, \}\,,$$

can be expressed by the identical formula when $g_0(a)$ is replaced by $g_L(\mu_L)$.

This requires that our choice of $Z_g^L$ satisfy:

$$Z_g^L(t = 0, g_L(\mu_L = \frac{1}{a})) = 1 \ . \qquad (18)$$

Eq.(18) insures that the renormalized coupling, $g_L(\mu_L)$ and the bare coupling, $g_0(a)$ agree for $\mu_L = 1/a \gg \Lambda_L$. By choosing the prescription as above, we have defined $\Lambda_L$ in the conventional way [1, 2, 4].

Having made a choice for $\mu_L$ as above, we know that the ratio, $\mu_L/\mu_{\overline{ms}}$ is equal to $\Lambda_L/\Lambda_{\overline{ms}}$ in order to keep the coupling common in both renormalization schemes. In this sense, a renormalization scale on the lattice, $\mu_L = 1/a \approx 2$ GeV (Eq.(4), Eq.(5) and Eq.(6)) corresponds to about $\mu_{\overline{ms}} \approx 90$ GeV [23] for the $\beta = 5.7$ numerical simulations [7, 8, 9].

The natural question is what choice of $\mu_L$ is best to do perturbation. Experiment suggests that the perturbative expansion converges most rapidly if we choose the renormalization scale equal to the energy-momentum of the actual physical process in the $\overline{MS}$ scheme rather than in other schemes ($\mu_{\overline{ms}} = \sqrt{s}$ [20]). Hence we know that $\mu_{\overline{ms}}$ = physical energy-momentum in the $\overline{MS}$ scheme is the best to do perturbation [20]. When working with this quark mass parameter, $M_0(a)$ initially defined at the lattice scale a, it is reasonable to use $\mu_{\overline{ms}} = 1/a$ to set the physical energy scale. So $\mu_{\overline{ms}} \approx 2$ GeV is chosen, which corresponds to $\mu_L \approx 50$ MeV on the lattice. The point is that it is better to choose $\mu_{\overline{ms}} = 1/a$ than $\mu_L = 1/a$ in order to improve the convergence of the perturbation series.

For the mass and wave function renormalization constants, we will leave our prescription quite flexible requiring only:

$$Z_m^L(t = C_1, g_L) = 1 \ , \quad Z_\psi^L(t = C_2, g_L) = 1 \qquad (19)$$

where the uncertainties, now transferred to $C_1$ and $C_2$ will be determined later in Section 5.

## 4.2 One-Loop Self-Energy and Mass Renormalization

The Feynmann rules and gauge-fixing(Feynman gauge) we use are consistent with Ref.[4, 5, 15]. Only the two Feynmann diagrams in Figure 1 can contribute to the one-loop self-energy, as follows:

$\Sigma(p, M_0(a), g_0, a)$

$$= \frac{g_0^2}{a} R(N_c) \sum_\mu \int_{-\pi}^{+\pi} \frac{d^4k}{(2\pi)^4} \gamma_\mu \frac{\frac{1}{2}(1 + \cos(2ap_\mu + k_\mu))}{\sum_\nu i\gamma_\nu \sin(ap_\nu + k_\nu) + aM_0} \gamma_\mu \frac{1}{4\sum_\alpha \sin^2(\frac{k_\alpha}{2})}$$

$$-\frac{g_0^2}{2a} R(N_c) \sum_\rho i\gamma_\rho \sin(ap_\rho) \int_{-\pi}^{+\pi} \frac{d^4k}{(2\pi)^4} \frac{1}{4\sum_\tau \sin^2(\frac{k_\tau}{2})}$$

$$= \sum_\mu i\gamma_\mu p_\mu \Sigma_1^L + M_0(a) \Sigma_2^L \quad (20)$$

Here $N_c$ is the number of colors and $R(N_c)$ is given in Section 4. The $\gamma_\mu$ are the normal $4 \times 4$ Euclidean Dirác matrices and the expression in Eq.(20) is independent of color and flavor. It is the same for $N_f = 4$ as the propagator derived in Ref.[5]. $\Sigma_1^L$ and $\Sigma_2^L$ in Eq.(20) are

$$\Sigma_1^L = \frac{g_0^2}{8\pi^2} R(N_c) \left( 8\pi^2 \tau - \ln(a\mu_L) - \int_0^1 dx(1-x) \ln\left[\frac{\Delta}{\mu_L^2}\right] \right) + O(a) \quad (21)$$

$$\Sigma_2^L = \frac{g_0^2}{8\pi^2} R(N_c) \left( 8\pi^2 \sigma - 4\ln(a\mu_L) - 2\int_0^1 dx \ln\left[\frac{\Delta}{\mu_L^2}\right] \right) + O(a) \quad (22)$$

and $\tau = -0.044566$, $\sigma = 0.19745$ [22].

In the above one-loop calculation, it is assumed that $1/a \gg p$. As in Eq.(11) and Eq.(12), the renormalized propagator, $G_L$ is related to the bare propagator, $G_0$, as follows:

$$G_L^{-1}(p, M_L, g_L, \mu_L) = Z_\psi^L G_0^{-1}(p, M_0, g_0, a) \vert_{M_0 = Z_m^L M_L, \; g_0 = Z_g^L g_L} \quad (23)$$

where

$$G_0^{-1}(p, M_0(a), g_0, a) = \sum_\mu i\gamma_\mu p_\mu + M_0(a) + \Sigma(p, M_0(a), g_0, a) \quad (24)$$

From the prescriptions in section 4.1, we can determine the renormalization constants up to $g_L^2$ order, as follows:

$$Z_\psi^L = 1 + \frac{g_L^2}{8\pi^2} R(N_c) \left( \ln(a\mu_L) + \frac{C_2}{2} \right)$$

$$Z_m^L = 1 + \frac{g_L^2}{8\pi^2} R(N_c) \left( 3\ln(a\mu_L) + \frac{3C_1}{2} \right), \quad (25)$$

so that the renormalized quark propagator for $N_c = 3$ is

$$G_L^{-1}(p, M_L, g_L, \mu_L) = \left\{ 1 + \frac{g_L^2}{6\pi^2} \left( 8\pi^2 \tau + C_2 - \int_0^1 dx(1-x) \ln\left[\frac{\Delta}{\mu_L^2}\right] \right) \right\}$$

$$\cdot \left( i\not{p} + M_L \left\{ 1 + \frac{g_L^2}{6\pi^2} \left( 8\pi^2(\sigma - \tau) + \frac{3C_1}{2} - \int_0^1 dx(1+x) \ln\left[\frac{\Delta}{\mu_L^2}\right] \right) \right\} \right)$$

If we suppose that the pole is located at $p^2 = -M_{phy}^2$, $M_{phy}$ should satisfy the following condition:

$$M_{phy} = M_L \left\{ 1 + \frac{g_L^2}{6\pi^2} \left( 8\pi^2(\sigma - \tau) + \frac{3C_1}{2} - \int_0^1 dx(1+x) \ln\left[\frac{\Delta}{\mu_L^2}\right] \right) \right\} \quad (26)$$

where $\Delta = \Delta(p^2 = -M_{phy}^2)$ .

Eq.(26) will be used later to relate $M_L$ and $M_{\overline{ms}}$ defined in the two different schemes.

## 5 Connection between the Lattice and Continuum Parameters

The bridge condition will be constructed, exploiting the similar form the renormalization group equations take in the two schemes through one-loop order and choosing the same pole structure for the quark parpagator in both schemes.

### 5.1 The Bridge Condition

In the framework of a mass-independent renormalization scheme, the renormalization group determines the renormalized coupling as a function of only $\mu/\Lambda$ and the renormalized mass as the product of the renormalization group invariant mass, $\bar{M}$ and a function of only $\mu/\Lambda$ [10], where $\Lambda$ is an integration constant chosen by standard convention. Then the following two conditions can be chosen:

- Coupling relation : The renormalized couplings are the same for the introduced renormalization scales. For example,

$$g_L(\mu_L) = g_{\overline{ms}}(\mu_{\overline{ms}}) = g \iff \frac{\mu_L}{\Lambda_L} = \frac{\mu_{\overline{ms}}}{\Lambda_{\overline{ms}}} \quad (27)$$

because using the conventional definition of the $\Lambda$ parameter in both schemes we have

$$g_L(\mu_L) = f(\frac{\mu_L}{\Lambda_L}) \text{ and } g_{\overline{ms}} = f(\frac{\mu_{\overline{ms}}}{\Lambda_{\overline{ms}}}) ,$$

for the same function $f()$ through two loops.

- Mass relation : The renormalization group invariant mass is common to any renormalization scheme that is independent of dimensionful lagrangian parameters. For example,

$$\frac{\mu_L}{\Lambda_L} = \frac{\mu_{\overline{ms}}}{\Lambda_{\overline{ms}}} \text{ and common } \bar{M} \iff M_L(\mu_L, \bar{M}) = M_{\overline{ms}}(\mu_{\overline{ms}}, \bar{M}) \quad (28)$$

again because

$$M_L(\mu_L, \bar{M}) = \bar{M} h(\frac{\mu_L}{\Lambda_L}) \text{ and } M_{\overline{ms}}(\mu_{\overline{ms}}, \bar{M}) = \bar{M} h(\frac{\mu_{\overline{ms}}}{\Lambda_{\overline{ms}}}) ,$$

through one-loop order.

In particular, for $\beta(g) = -\beta_0 g^3 - \beta_1 g^5 - \beta_2 g^7 - \cdots$, it is a well-known fact that the first two coefficients ($\beta_0$ and $\beta_1$) are independent of the renormalization scheme while all the other coefficients depend on the renormalization scheme. However for $\gamma_m(g) = \gamma_0 g^2 + \gamma_1 g^4 + \cdots$, only the first coefficient $\gamma_0$ is independent of the renormalization scheme[20]. In the framework of the bridge condition, only the scheme-independent parts of the $\beta$ and $\gamma_m$ functions are considered and the scheme-dependent parts ignored so that $f(\frac{\mu}{\Lambda})$ and $h(\frac{\mu}{\Lambda})$ are universal i.e. independent of the renormalization scheme up to the order of our present calculation. The constant $C_1$ introduced in Eq.(19) will be determined only up to one-loop order. $\frac{\Lambda_L}{\Lambda_{\overline{ms}}}$ is estimated up to one-loop order[1, 2, 4] and has no higher order corrections[3].

The location of the pole in the two-point Green's function should be independent of the renormalization scheme[17] and determines the physical mass of the particle. The location of the pole in the quark propagator is independent of the renormalization scheme. The constant $C_1$ can be chosen explicitly to conform to the conventions of Eq.(28), by demanding that the pole location in the quark propagator is the same for both $\overline{MS}$ and lattice renormalization scheme.

Therefore from Eq.(14), Eq.(26), Eq.(27) and Eq.(28),

$$\frac{M_L(\mu_L, \bar{M})}{M_{\overline{ms}}(\mu_{\overline{ms}}, \bar{M})} = 1 - \frac{g_{\overline{ms}}^2}{6\pi^2} \left( 8\pi^2(\sigma - \tau) + \frac{1}{2} + \frac{3C_1}{2} + 3 \ln\left[\frac{\mu_L}{\mu_{\overline{ms}}}\right] \right) = 1 ,$$

$$\text{so } C_1 = \frac{16\pi^2}{3}(\tau - \sigma) - \frac{1}{3} + 2 \ln\left[\frac{\Lambda_{\overline{ms}}}{\Lambda_L}\right] , \quad (29)$$

where $\tau$ and $\sigma$ are given in Eq.(22). Various values of $C_1$ for $N_f = 0, 2, 3, 4, 5, 6$ and $8$, are given in Table 1.

## 5.2 Renormalized Coupling Constant

There are a number of ways to obtain the renormalized coupling constant, $g(\mu)$. Once we know the QCD $\Lambda$ parameter, we know the coupling constant $g(\mu)$. For example, one may use the experimental value of the $\Lambda$ parameter in the $\overline{MS}$ scheme [31]. For a given renormalization scale $\mu_{\overline{ms}}$, we can obtain the renormalized coupling constant, $g_L(\mu_L)$ since the ratio, $\Lambda_L/\Lambda_{\overline{MS}}$ is well-known [1, 2, 3, 4]. Another method uses the bare coupling on the lattice in order to obtain the renormalized coupling. In this article, the latter method is chosen in order to do everything consistently in terms of the lattice parameters and the lattice observables.

From Eq.(15), the renormalization group equation can be derived, as follows:

$$\left(-2\frac{\partial}{\partial t} + \beta(g_L)\frac{\partial}{\partial g_L}\right) g_L Z_g^L(t, g_L) = 0 \tag{30}$$

where $\beta(g) = -\beta_0 g^3 - \beta_1 g^5 - ....$

and $\beta_0 = \dfrac{11}{16\pi^2}(1 - \dfrac{2N_f}{33})$, $\beta_1 = \dfrac{1}{256\pi^4}(102 - \dfrac{38N_f}{3})$

Eq.(30) can be solved, with Eq.(18) as a boundary condition [6], up to the leading order in $g_L^2$,

$$\begin{aligned} Z_g^L(t, g_L) &= \frac{1}{\sqrt{1 + t\beta_0 g_L^2}} \\ g_{\overline{ms}}^2(\mu_{\overline{ms}}) &= g_L^2(\mu_L) = \frac{g_0^2(a)}{1 - t\beta_0 g_0^2(a)} \end{aligned} \tag{31}$$

where

$$t = -2\ln(a\mu_L) = -2\left[\ln(a\mu_{\overline{ms}}) + \ln\left(\frac{\Lambda_L}{\Lambda_{\overline{ms}}}\right)\right] \tag{32}$$

From Eq.(31), for various $N_f$ and $\mu_{\overline{ms}}$, $g_{\overline{ms}}$ can be obtained.

## 5.3 Renormalized Mass

We have already obtained an expression for the mass renormalization constant, $Z_m^L$ in Eq.(25). Thus the naive relationship between the renormalized quark mass in $\overline{MS}$ and the bare quark mass in the lattice formulation follows from Eq.(16), Eq.(25), Eq.(27), Eq.(28) and Eq.(29):

$$M_{\overline{ms}}(\mu_{\overline{ms}}) = M_L(\mu_L) = \frac{1}{Z_m^L} M_0(a)$$

$$M_{\overline{ms}}(\mu_{\overline{ms}}) = \left[1 - \frac{g_L^2}{2\pi^2}\left(\ln(a\mu_{\overline{ms}}) + \ln\left(\frac{\Lambda_L}{\Lambda_{\overline{ms}}}\right) + \frac{C_1}{2}\right)\right] M_0(a)$$

$$M_{\overline{ms}}(\mu_{\overline{ms}}) = \left[1 - \frac{g_{\overline{ms}}^2}{2\pi^2}\left(\ln(a\mu_{\overline{ms}}) + \frac{8\pi^2}{3}(\tau - \sigma) - \frac{1}{6}\right)\right] M_0(a) \quad (33)$$

As described in section 4.1, it is reasonable to use $\mu_{\overline{ms}} = 1/a$ as condition for the optimal perturbative expansion. From Eq.(33),

$$M_{\overline{ms}}(1/a) = \left[1 - \frac{g_{\overline{ms}}^2}{2\pi^2}\left(\frac{8\pi^2}{3}(\tau - \sigma) - \frac{1}{6}\right)\right] M_0(a) \quad (34)$$

Solving the renormalization group equation, we can obtain the renormalized quark mass at any other scale in the following way.

$$\gamma_m(g) = -\mu_{\overline{ms}} \frac{\partial \ln(M_{\overline{ms}})}{\partial \mu_{\overline{ms}}} = \gamma_0 g^2 + \gamma_1 g^4 + \ldots \quad (35)$$

$$\text{where } \gamma_0 = \frac{1}{2\pi^2}$$

We can solve Eq.(35) such that the renormalization scale runs from $1/a$ to an arbitrary scale, $\mu_{\overline{ms}}$. Then the solution to Eq.(35) is

$$M_{\overline{ms}}(\mu_{\overline{ms}}) = \left(\frac{g_{\overline{ms}}^2(\mu_{\overline{ms}})}{g_{\overline{ms}}^2(1/a)}\right)^{\frac{\gamma_0}{2\beta_0}} M_{\overline{ms}}(1/a)$$

$$= \left(\frac{g_{\overline{ms}}^2(\mu_{\overline{ms}})}{g_{\overline{ms}}^2(1/a)}\right)^{\frac{\gamma_0}{2\beta_0}} \left[1 - \frac{g_{\overline{ms}}^2}{2\pi^2}\left(\frac{8\pi^2}{3}(\tau - \sigma) - \frac{1}{6}\right)\right] M_0(a) \quad (36)$$

Other authors quote values for $M_{\overline{ms}}(\mu_{\overline{ms}})$ evaluated at $\mu_{\overline{ms}} = 1$ GeV. From Eq.(36), we can obtain the renormalized quark mass at that 1 GeV scale in the $\overline{MS}$ scheme(Table 1), although we are entering a region where perturbation is unlikely to be accurate.

Equation (33) is just the lowest order relationship and can be improved through leading logarithmic approximation. From Eq.(16), the renormalization group equation is derived as follows:

$$\left(-2\frac{\partial}{\partial t} + \beta(g_L)\frac{\partial}{\partial g_L} - \gamma_m(g_L)\right) Z_m^L(t, g_L) = 0 \qquad (37)$$

The solution of Eq.(34) with Eq.(19) as a boundary condition is, up to the leading order in $g_L^2$ [6],

$$Z_m^L(t, g_L) = \left[1 + \beta_0 g_L^2(t - C_1)\right]^{-\frac{\gamma_0}{2\beta_0}}$$

and from Eq.(16), Eq.(27) and Eq.(28),

$$M_{\overline{ms}}^{L.L.A.}(\mu_{\overline{ms}}) = \frac{1}{Z_m^L} M_0(a) = [1 + \beta_0 g_{\overline{ms}}^2(t - C_1)]^{\frac{\gamma_0}{2\beta_0}} M_0(a) \qquad (38)$$

where the renormalized coupling and $t$ can be obtained from Eq.(31) and Eq.(32). As one can see in the above, up to $g_{\overline{ms}}^2$ order, Eq.(38) is identical to Eq.(33). From Eq.(36) and Eq.(38), for various $N_f$, for $\beta = 5.7$ and for $\mu_{\overline{ms}} = 1$ GeV, $M_{\overline{ms}}$ and $M_{\overline{ms}}^{L.L.A.}$ are given in Table 1.

## 5.4 Renormalization Group Invariant Mass

Using the conventions of Ref.[10] and expanding the function, $\gamma_m(g)$, the renormalization group invariant mass is defined, up to leading order in $g_L^2$, by

$$\bar{M} = \frac{M_L}{(\beta_0 g_L^2)^{\frac{\gamma_0}{2\beta_0}}} . \qquad (39)$$

From Eq.(15), Eq.(16), Eq.(31), Eq.(38) and Eq.(39), $\bar{M}$ is related to $M_0(a)$, as follows:

$$\bar{M} = \left(\frac{1}{\beta_0 g_0^2(a)} - C_1\right)^{\frac{\gamma_0}{2\beta_0}} M_0(a) \qquad (40)$$

Eq.(40) is independent of the renormalization scale, $\mu_L$ or $\mu_{\overline{ms}}$, which means that $\bar{M}$ is less sensitive to the renormalization scheme. It is reasonable to choose the renormalization group invariant mass as the physical quantity to extract from a lattice calculation, since we don't need to introduce the renormalization scale, $\mu_L$ or $\mu_{\overline{ms}}$. The renormalization group invariant quark mass is given in Table 1, from Eq.(40).

## 5.5 Light Quark Mass for $N_f = 2$, $\beta = 5.7$

In this section, it is explicitly explained how the renormalized mass at the given scale (1 GeV in $\overline{MS}$ scheme) and the renormalization group invariant mass[10] can be estimated in the manner described above using the numerical results of the lattice QCD simulation.

Let us start with the renormalized coupling, $g_{\overline{ms}}(\mu_{\overline{ms}})$ or $g_L(\mu_L)$. We know that $\beta = 5.7$ corresponds to the bare coupling constant, $g_0^2(a) = 1.05263$ since $g_0^2(a) = 6/\beta$. From Ref.[4] and Table 1, it is known that $\Lambda_L/\Lambda_{\overline{ms}} = 0.02279$ for $N_f = 2$. Since $\mu_L = (\Lambda_L/\Lambda_{\overline{ms}})\mu_{\overline{ms}}$, we know that $\mu_{\overline{ms}} = 1$ GeV corresponds to $\mu_L = 0.02279$ GeV. From Eq.(4), Eq.(5), Eq(6) and Ref.[7], $a^{-1} = 2.23$ GeV and $t = -2\ln(a\mu_L) = 9.1707$. For $N_f = 2$, $\beta_0 = \frac{29}{48\pi^2}$, which is defined in Eq.(30). Since we know $g_0(a)$, $t$ and $\beta_0$, the renormalized coupling constant at a scale of 1 GeV is from Eq.(31),

$$g_{\overline{ms}}^2(\mu_{\overline{ms}} = 1 GeV) = g_L^2(\mu_L = 23 MeV) = 2.57 \ . \tag{41}$$

In a similar way, we can obtain the renormalized coupling constant at a scale of $1/a$:

$$g_{\overline{ms}}^2(\mu_{\overline{ms}} = 1/a) = g_L^2(\mu_L = 51\ MeV) = 2.05 \ . \tag{42}$$

Now let us go ahead to obtain the renormalized light quark mass at a scale of $1/a$. We know $g_{\overline{ms}}(1/a)$ and $(\tau - \sigma)$ from Eq.(22). The bare quark mass $M_0(a) = 1.6$ MeV at most from Eq.(4), Eq.(5) and Ref.[7, 9]. So the renormalized light quark mass at a scale of $1/a$ is naively, from Eq.(34),

$$\frac{M_{\overline{ms}}(1/a)}{M_0(a)} = 1.68$$
$$M_{\overline{ms}}(1/a) = M_L(51\ MeV) = 2.69 MeV. \tag{43}$$

We can also obtain the renormalized quark mass at a scale of 1 GeV from Eq.(41), Eq.(42) and Eq.(36).

$$\frac{M_{\overline{ms}}(1\ GeV)}{M_0(a)} = 1.84$$
$$M_{\overline{ms}}(1\ GeV) = M_L(51\ MeV) = 2.95\ MeV\ . \qquad (44)$$

The renormalized light quark mass at a scale of 1 GeV, improved by leading logarithmic approximation is, from Eq.(38),

$$M_{\overline{ms}}^{L.L.A.}(1\ GeV) = M_L^{L.L.A.}(51\ MeV) = 1.64\ M_0(a) = 2.62\ MeV\ . \qquad (45)$$

Now let us obtain the renormalization group invariant mass. Knowing $\Lambda_L/\Lambda_{\overline{ms}} = 0.02279$ from Table 1 and Ref.[4], we can determine $C_1 = -5.5098$ from Eq.(29). Since we know $\gamma_0$, $\beta_0$, $g_0(a)$ and $C_1$, the renormalization group invariant mass is, from Eq.(40),

$$\bar{M} = 3.52 M_0(a) = 5.63\ MeV\ . \qquad (46)$$

Eq.(33) is just a crude result of one-loop renormalization, while Eq.(40) is more reliable since it includes all the leading logarithmic contributions and is independent of all the renormalization scales introduced, which means less dependence on the conventions. The 10% difference between the numbers in Eq.(44) and Eq.(45) suggests the size of the omitted higher order corrections.

# 6  Mean Field Theory

Lepage and Mackenzie showed that tadpole diagrams are the main source of the large difference between the bare lattice coupling, $g_0(a)$ and the renormalized coupling, $g_{\overline{MS}}(\mu_{\overline{MS}} = \frac{1}{a})$ [12]. They suggest a meanfield method for removing the dominant effect of tadpole diagrams. Here we apply their meanfield method to the staggered fermion formalism, in order to improve the estimation of the renormalized quark mass on the lattice non-perturbatively.

## 6.1  Tadpole Improvement for the Coupling Constant

The matching of lattice operators with continuum operators is based on the expansion

$$U_\mu(x) \equiv e^{iagA_\mu(x)} \rightarrow 1 + iagA_\mu(x) \qquad (47)$$

when the lattice spacing a is small. But higher order terms in the expansion of $U_\mu$ contain addtional factors of $agA_\mu$ and the contraction of $A_\mu(x)$'s with each other generates ultra-violet divergence (i.e. $\propto \frac{1}{a^n}$) which cancel the addional powers of a (i.e. $\propto a^n$). These terms are of the order of $g^n$, not suppressed by the powers of lattice spacing a, and are very large. These are called the QCD tadpole contributions[12].

These large tadpole contribution causes poor perturbative expansion in lattice QCD. Lepage and Mackenzie suggested the mean field method[12] in order to refine naive perturbative expansion by removing tadpole contribution. They noticed that the vacuum expectation value of the link matrix is smaller than 1. They suggested that the appropriate connection with the continuum gauge field is

$$U_\mu(x) \to u_0(1 + iagA_\mu(x)) \tag{48}$$

where $u_0$ represents the mean value of the link. They choose $u_0$ in a gauge invariant way:

$$u_0 \equiv \left[\text{Re} < \frac{1}{3}\text{Tr}U_\square >\right]^{\frac{1}{4}} \tag{49}$$

Other choices of $u_0$ are possible and equally good. It is the rescaled links $\frac{U_\mu}{u_0}$ that should be expanded around unity. The staggered fermion action given in Eq.(1) can be rewritten in terms of $\psi \equiv \sqrt{u_0}\chi$

$$S = -\sum_{x,\mu}\frac{1}{2}\eta_\mu(x)[\bar{\psi}(x)\frac{U_\mu^\dagger(x)}{u_0}\psi(x+a_\mu) - \bar{\psi}(x+a_\mu)\frac{U_\mu(x)}{u_0}\psi(x)]$$
$$-\frac{M}{u_0}a\sum_x \bar{\psi}(x)\psi(x) + S_{gluon} \tag{50}$$

$$S_{gluon} = -\sum_\square \beta_{MF}\frac{1}{N_c}\frac{\text{Re}\text{Tr}U_\square}{u_0^4} \tag{51}$$

where

$$\beta_{MF} = \frac{2N_c}{g_{MF}^2}, \ g_{MF}^2 = \frac{g_0^2(a)}{\text{Re} < \frac{1}{3}\text{Tr}U_\square >} \tag{52}$$

The perturbative expansion for $\text{Tr}U_\square$ given in Ref.[32] is

$$\text{Re} < \frac{1}{3}\text{Tr}U_\square >= 1 - \frac{1}{3}g_0^2(a) + O(g_0^4(a)) \tag{53}$$

We can relate $g_{MF}^2$ to $g_0^2$ perturbatively as follows:

$$g_{MF}^2 = \frac{g_0^2(a)}{\text{Re} < \frac{1}{3}\text{Tr}U_\Box >} = g_0^2(a)[1 + \frac{1}{3}g_0^2(a) + O(g_0^4(a))] \tag{54}$$

From Eq.(31), Eq.(32) and Eq.(54) we can also relate $g_{MF}^2$ to $g_{\overline{ms}}^2(\mu_{\overline{ms}})$ perturbatively:

$$g_{\overline{ms}}^2(\mu_{\overline{ms}}) = g_{MF}^2[1 - \beta_0 g_{MF}^2\{2\ln(a\mu_{\overline{ms}}) + 2\ln\left(\frac{\Lambda_L}{\Lambda_{\overline{ms}}}\right)\}$$
$$-\frac{1}{3}g_{MF}^2 + O(g_{MF}^4)] \tag{55}$$

$$\text{for } N_f = 2, \quad g_{\overline{ms}}^2(\mu_{\overline{ms}}) = g_{MF}^2[1 - \beta_0 g_{MF}^2\{2\ln(a\mu_{\overline{ms}}) - 2.117\}$$
$$+O(g_{MF}^4)] \tag{56}$$

$$\text{for } \mu_{\overline{ms}} = \frac{\pi}{a}, \quad g_{\overline{ms}}^2(\frac{\pi}{a}) = g_{MF}^2[1 - 0.0106 g_{MF}^2 + O(g_{MF}^4)] \tag{57}$$

As one can see in Eq.(57), $g_{MF}^2$ is extremely close to $g_{\overline{ms}}^2(\frac{\pi}{a})$, where the difference between these two coupling constants is only a few percent from the standpoint of pertubative expansion. This means that by running the renormalized coupling in the $\overline{MS}$ scheme at a scale of $\frac{\pi}{a}$, one can absorb the large tadpole contribution into the renormalized coupling constant.

## 6.2 Tadpole Improvement for the Quark Mass

As the lattice spacing, $a$ goes to zero, the action given in Eq.(50) will be

$$S = -\int_x \bar{\psi}(x)[D \cdot \gamma + \frac{M}{u_0}]\psi(x) + S_{gluon} \tag{58}$$

Here the claim is that $\psi$ is better matched to the continuum quark field than $\chi$. From the above action, we notice that $M_{MF} \equiv \frac{M_0(a)}{u_0}$ takes the place of the bare quark mass on the lattice.

Hence the renormalized quark mass can be obtained as follows:

$$\begin{aligned} M_{\overline{ms}}(\mu_{\overline{ms}}) &= Z_m(a\mu_{\overline{ms}}, g_{\overline{ms}}^2(\mu_{\overline{ms}}))M_0(a) \\ &= Z_m u_0 M_{MF}(a) \\ &\equiv \tilde{Z}_m((a\mu_{\overline{ms}}, g_{MF}^2)M_{MF}(a) \end{aligned} \tag{59}$$

where $\tilde{Z}_m \equiv Z_m u_0$.

From Eq.(33), Eq.(49), Eq.(53), Eq.(54) and Eq.(55), the perturbative expansion of $\tilde{Z}_m$ can be obtained as follows:

$$\tilde{Z}_m((a\mu_{\overline{ms}}, g^2_{MF})$$
$$= Z_m(a\mu_{\overline{ms}}, g^2_{\overline{ms}}(\mu_{\overline{ms}}))u_0$$
$$= \left[1 - \frac{g^2_{\overline{ms}}}{2\pi^2}\left(\ln(a\mu_{\overline{ms}}) + \frac{8\pi^2}{3}(\tau - \sigma) - \frac{1}{6}\right)\right]_{g^2_{\overline{ms}} = g^2_{MF} + O(g^4_{MF})}$$
$$\cdot \left[1 - \frac{1}{12}g^2_{MF}\right]$$
$$= \left[1 - \frac{g^2_{MF}}{2\pi^2}\left(\ln(a\mu_{\overline{ms}}) + \frac{8\pi^2}{3}(\tau - \sigma) - \frac{1}{6} + \frac{\pi^2}{6}\right)\right] \quad (60)$$

The numerical data for $u_0$ is obtained by the Columbia lattice group at $\beta = 5.7$ for $N_f = 2$ [9]:

$$u_0^4 = \text{Re} < \frac{1}{3}\text{Tr}U_\square >= 0.58 \quad (61)$$
$$u_0 = 0.87 \quad . \quad (62)$$

Putting everything altogether, we can obtain the renormalized quark mass:

$$M_{MF}(a) = 1.83 MeV \quad (63)$$
$$\tilde{Z}_m(1, g^2_{MF} = 1.815) = 1.45 \quad (64)$$
$$M_{\overline{ms}}(\frac{1}{a}) = 2.66 MeV \quad (65)$$

This final result (Eq.(65)) is in good agreement with that of Eq.(43) in the section 5.5. This good agreement means that in fact through perturbative expansion with respect to $g_{\overline{ms}}(1/a)$, one can remove the tadpole contributions, which is very large when one does the pertubative expansion with respect to the lattice bare coupling, $g_0^2$.

# 7  Light Quark Mass from QCD sum rule

There are various methods to determine the light quark mass in continuum QCD[10, 25, 26, 27]. The actual determination of the light quark mass in the framework of QCD can be done reasonably through the QCD sum rule formalism by Shifman, Voloshin and Zakharov (S.V.Z.) [28, 29]. In that formalism, they study the two point correlation function of the divergence of the axial current (isospin sector). They determine the running quark mass at 1.2 GeV scale with $N_f = 3$ i.e. three dynamical quarks moving around. In the $\overline{MS}$ scheme, they choose renormalization scale $\mu_{\overline{ms}}$ to be the same as the energy-momentum scale of the physical process. Their choice, $\mu_{\overline{ms}} = 1 \sim 1.2$ GeV means physically that the charm and bottom quarks are decoupled[30] and that only three light quarks contribute to the dynamics of the QCD. In other words, at $\mu_{\overline{ms}} = 1 \sim 1.2$ GeV, we can not decouple the strange quark from the dynamics of the QCD. Now we have a problem that lattice QCD is simulated numerically with 2 dynamical quarks but the QCD sum rule assumes 3 dynamical quarks.

From the recent experimental data[31], we have the following QCD $\Lambda$ parameter:

$$\Lambda_{\overline{ms}}(N_f = 3) \cong 250 \pm 150 MeV$$

where the error is partly a combination of statistial and sytematic errors and partly due to the scale uncertainty. We should use $\Lambda_{\overline{ms}}(N_f = 3)$ at the scale below the charm quark mass. Using $\Lambda_{\overline{ms}}(N_f = 3) \cong 250 \pm 150$ MeV, the QCD sum rule gives the following light quark mass[10]. For $N_f = 3$,

$$M_{\overline{ms}}(\mu_{\overline{ms}} = 1 GeV) = \frac{M_u + M_d}{2} = 3.5 \sim 9.0 \text{ MeV}$$
$$\bar{M} = 4.0 \sim 13.0 \text{ MeV} ,$$

where the larger value corresponds to the smaller $\Lambda_{\overline{ms}} < 150$ MeV and the smaller value corresponds to the larger $\Lambda_{\overline{ms}} > 300$ MeV.

The QCD sum rules take into consideration the resonance contribution but not the continuum contribution to the imaginary part of the two point correlation function of the axial current divergence. Omitting this continuum contribution from the non-perturbative low-energy region can introduce around 20% errors to the above expectation value of the light quark mass[28].

Even though there is a difference in the number of dynamical quarks, the lattice QCD expectation value of the quark mass with $N_f = 2$ given in Section 6.5 agrees with that of the S.V.Z. QCD sum rule with $N_f = 3$ in the above.

But in order to do the exact comparison, the lattice QCD simulation with three dynmical quarks (three sea quarks: one of them is 10 ∼ 20 times heavier than the other two light sea quarks) is necessary since the strange quark can not be decoupled from the QCD dynamics at the energy-momentum scale of 1 GeV[30].

# 8  Conclusion

The QCD dynamical quark mass is renormalized in two different renormalization schemes ($\overline{MS}$ and lattice-regularized minimal subtraction). The bridge conditions are chosen to make connection between the two schemes. The ratios of the continuum renormalized quark mass to the lattice bare quark mass are given in Table 1.

Ref.[7] contains an earlier attempt to compute the renormalized quark mass. Their value of 2.2 MeV is somewhat different from that of this article(2.95 MeV). Fukugita et al. use the one-loop renormalization and Eq.(3.7) in Ref.[7] is the same as Eq.(33) except for the coupling constant that appears. This 25% difference comes from the different choice of the coupling: the bare coupling on the lattice is used in Ref.[7] while the renormalized coupling is used in this article. It has been pointed out that the lattice bare coupling constant may be a poor choice as an expansion parameter and that the use of the improved coupling constant including renormalization due to gluon tadpole contributions improves the reliability of the lattice perturbative expansion[12]. Using the mean field method suggested by Lepage and Mackenzie, the tadpole-improved renormalized quark mass is obtained, which is extremely close to that obtained by the use of the renormalized coupling in the $\overline{MS}$ scheme at $\frac{1}{a}$ scale. It has been proposed in other contexts [20] to use the renormalized coupling (in the $\overline{MS}$ scheme with the renormalization scale equal to the physical energy-momentum) instead of the bare coupling. At the least, the differences between the results obtained here and those of Fukugita et al. represent the size of the perturbative errors.

The QCD sum rule predicts that the renormalized quark mass (at a 1 GeV scale in the $\overline{MS}$ scheme) is 3.5 ∼ 9.0 MeV and that the renormalization group invariant mass is 4.0 ∼ 13.0 MeV for $N_f = 3$, with 20% uncertainty [10, 28, 29]. The renormalized quark mass (at a 1 GeV scale in the $\overline{MS}$ scheme) obtained from the lattice QCD simulation is 2.95 MeV and the renormalization group invariant mass 5.63 MeV for $N_f = 2$.

The difference between 2-flavor dynamics and 3-flavor dynamics with one of the three flavors much heavier may be presumed to be so small that the comparison between 2-flavor and 3-flavor dynamics may make physical sense.

There are two other sources of error in the lattice QCD simulation: one is finite-temperature effect and the other finite-volume effect. In the hadron mass calculations on $16^3 \times 16$, $16^3 \times 32$ and $32^3 \times 32$ shows approximately $5 \sim 10\%$ effect[33]. These small effects are completely neglected in this article. But the systematic analysis of finite size effects should be done to look into physics on the lattice more precisely.

# 9  Acknowledgements

I am indebted a lot to Prof. Norman H. Christ. This work could not have been done without his consistent help and encouragement. Helpful conversations with Prof. R. Friedberg, Prof. Robert D. Mawhinney, Prof. A. Mueller and Prof. V.P. Nair are acknowledged with gratitude.

| $N_f$ | $\frac{\Lambda_L}{\Lambda_{\overline{ms}}}$ | $C_1$ | $\frac{M_{\overline{ms}}(1/a)}{M_0}$ | $\frac{M_{\overline{ms}}(1GeV)}{M_0}$ | $\frac{M_{\overline{ms}}^{L.L.A}(1GeV)}{M_0}$ | $\frac{\bar{M}}{M_0}$ |
|---|---|---|---|---|---|---|
| 0 | 0.0347 | -6.35 | 1.69 | 1.86 | 1.62 | 2.97 |
| 2 | 0.0228 | -5.51 | 1.68 | 1.84 | 1.64 | 3.52 |
| 3 | 0.0176 | -5.00 | 1.68 | 1.84 | 1.65 | 3.92 |
| 4 | 0.0131 | -4.40 | 1.67 | 1.83 | 1.67 | 4.45 |
| 5 | 0.00922 | -3.70 | 1.67 | 1.83 | 1.68 | 5.17 |
| 6 | 0.00608 | -2.87 | 1.67 | 1.82 | 1.70 | 6.19 |
| 8 | 0.00197 | -0.61 | 1.66 | 1.81 | 1.74 | 10.26 |

Table 1: Here we use $\beta = 5.7$, $a^{-1} = 2.23 GeV$. The ratio, $\Lambda_L/\Lambda_{\overline{ms}}$ is obtained from Ref.[4].

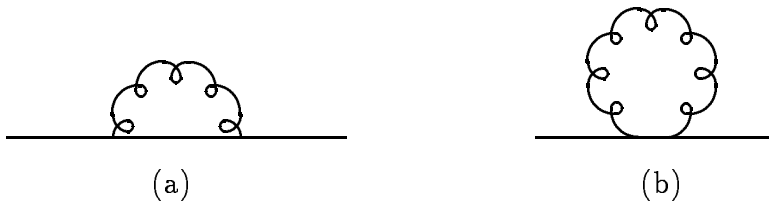

Figure 1: One-Loop Contribution to the Self-Energy (a) gluon exchage diagram (b) gluon bubble diagram